\documentclass[twocolumn,prb,showpacs,floatfix]{revtex4}
\usepackage{graphicx}
\usepackage{bm}

\newcommand{\be}{\begin{equation}}
\newcommand{\ee}{\end{equation}}
\newcommand{\bea}{\begin{eqnarray}}
\newcommand{\eea}{\end{eqnarray}}

\begin{document}

\title{\textbf{Temperature Dependence of the Magnetic Susceptibility for
Triangular-Lattice Antiferromagnets with spatially anisotropic
exchange constants}}

\author{
        Weihong Zheng$^{1}$,
        Rajiv R.~P.~Singh$^{2}$,
        Ross H. McKenzie$^{3}$, and
        Radu Coldea$^{4}$
}

\affiliation{ $^1$School of Physics, University of New South
Wales, Sydney NSW
2052, Australia\\
$^2$Department of Physics, University of California, Davis, CA
95616\\
$^3$Department of Physics, University of Queensland, Brisbane,
Australia\\
$^4$Department of Physics, Oxford University, Oxford OX1 3PU,
United Kingdom
\\
}

\date{\today}


\begin{abstract}
We present the temperature dependence of the
 uniform susceptibility of spin-half quantum antiferromagnets
on spatially anisotropic triangular-lattices, using high temperature
series expansions. We consider a model with two exchange constants,
$J_1$ and $J_2$ on a lattice that interpolates between the limits
of a square-lattice ($J_1=0$), a triangular-lattice ($J_2=J_1$), and
decoupled linear chains ($J_2=0$). In all cases, the susceptibility which has
a Curie-Weiss behavior at high temperatures, rolls over and begins to
decrease below a peak temperature, $T_p$. Scaling the exchange
constants to get the same peak temperature, shows that the susceptibilities
for the square-lattice and linear chain limits have similar
magnitudes near the peak.
Maximum deviation arises near the triangular-lattice limit,
where frustration leads to much smaller susceptibility and with
a flatter temperature dependence.
We compare our results to the inorganic materials Cs$_2$CuCl$_4$ and
Cs$_2$CuBr$_4$ and to a number of organic molecular crystals. We find
that the former (Cs$_2$CuCl$_4$ and Cs$_2$CuBr$_4$)
are weakly frustrated and their exchange parameters
determined through the temperature dependence of the susceptibility
are in agreement with neutron-scattering measurements.
In contrast, the organic materials considered are strongly frustrated
with exchange parameters near the isotropic triangular-lattice limit.
\end{abstract}

\maketitle

\section{Introduction}

Understanding the interplay of quantum and thermal fluctuations
and geometrical frustration in low-dimensional quantum
antiferromagnets is a considerable theoretical
challenge.\cite{aeppli,senthil,misguich2,greedan,harrison,ramirez}
 Research in frustrated
quantum antiferromagnets was greatly stimulated by Anderson's
``resonating valence bond'' (RVB) paper\cite{rvb} in which he
suggested that the parent insulators of the cuprate
superconductors might have spin liquid ground states and
excitations with fractional quantum numbers, motivated by his
earlier suggestion of such a ground state for the Heisenberg
antiferromagnet on the triangular lattice.\cite{anderson}
 The Ising model on a triangular lattice
illustrates the rich physics that can arise due to frustration:
 it is known to have
a macroscopic number of degenerate ground states.\cite{wannier}
The antiferromagnetic Heisenberg model
with spatially anisotropic exchange interactions
on the triangular lattice is of interest both
 theoretically and experimentally.
It describes the spin excitations in
 Cs$_2$CuCl$_4$\cite{coldea2}
and Cs$_2$CuBr$_4$\cite{ono} and the Mott insulating phase of
several classes of superconducting organic molecular
crystals\cite{mckenzie}. Other materials for which this model is
relevant include NaTiO$_2$,\cite{natio}
 CuCl$_2$ graphite intercalation compounds,\cite{suzuki} and
the anhydrous alum, KTi(SO$_4$)$_2$.\cite{bramwell}
Theoretically, this Heisenberg model
is a candidate for a system with
spin liquid ground states and possibly excitations with fractional
quantum numbers.\cite{anderson,laughlin,sondhi} For the triangular-lattice
model with spatially isotropic interactions, the preponderence
of numerical evidence \cite{lhuillier,huse,farnell,sorella}
 suggests that the ground
state has long-range magnetic order. However, making the
interactions spatially anisotropic can lead to a very rich ground-state
phase diagram \cite{weihong}.

The spatially anisotropic model, defined by a nearest-neighbor
exchange constant $J_1$ along one axis and $J_2$ along all other
axes [see Fig.\ \ref{fig1}], interpolates between the limits of
square-lattice ($J_1=0$), triangular-lattice ($J_2=J_1$) and
decoupled linear chain ($J_2=0$) limits.\cite{row,stephenson}
 It has been studied by
spin wave theory,\cite{merino}
 series expansions,\cite{weihong} large-N techniques,\cite{chung}
slave fermions,\cite{wen}
 Schwinger bosons with gaussian fluctuations,\cite{ceccatto}  and
variational quantum Monte Carlo.\cite{yunoki}
Quantum fluctuations are largest for $J_1 \simeq 0.8J_2$ and
for $J_1 > 4 J_2$,\cite{merino,weihong} and so for these parameter regions one
is most likely to observe quantum disordered phases.

From an experimental point of view,
it is highly desirable to have a definitive way to determine
the values of the exchange parameters for individual material
systems. Recently, it has been shown how for materials with
relatively small values for the Heisenberg
exchange constants $J$ this can be done in high magnetic fields,
using inelastic neutron scattering to measure the spin wave
dispersion in the field induced ferromagnetic phase.
\cite{coldea4}
The temperature dependence of the magnetic susceptibility is one
of the most common experimental measurements and it would be very
useful if that can be used to determine the extent of frustration
and the various exchange constants directly. It is particularly important
to have a scheme for materials, where the very high temperature
behavior of the system ($T\gg J$) is not accesible to experiments.
Previously, Castilla, Chakravarty, and Emery  pointed out how the
temperature dependence of the magnetic susceptibility of
the antiferromagnetic spin chain compound
CuGeO$_3$ implied significant magnetic frustration.\cite{emery} In that case,
it constrains the ratio of the nearest and next-nearest neighbour exchanges
along the chain.\cite{fabricius} Similarly, it is reasonable to expect that
the temperature dependence of the magnetic susceptibility should depend on
frustration in two-dimensional models also and hence constrain the
ratio $J_1/J_2$.

The Mott insulating phase of the organic molecular crystals
is of particular interest because under pressure
the materials considered become superconducting.
A possible RVB theory of superconductivity
in such materials, emphasising
the role of frustration, have recently been proposed.\cite{powell}
These materials
have exchange constants in the range of
several hundred Kelvin, and their behaviour has led to several puzzles.
Tamura and Kato\cite{tamura}
measured the temperature dependence of the magnetic
susceptibility for five organic molecular crystals
in the family,
 $\beta'$-[Pd(dmit)$_2$]X
(where dmit is the electron acceptor molecule
thiol-2-thione-4, 5-dithiolate, C$_3$S$_5$)
and the cation
X = Me$_4$As, Me$_4$P, Me$_4$Sb, Et$_2$Me$_2$P, and Et$_2$Me$_2$Sb,
 where Me = CH$_3$ and Et = C$_2$H$_5$, denote methyl and
 ethyl groups, respectively).
They compared their results with the predictions
for the square and triangular lattices and
found that for all the materials
the results could be fitted by the
high temperature series expansion
for the triangular lattice. However, some and not all of them
undergo a transition to a magnetically ordered state at
low temperatures.

Recently, Shimizu {\it et al.},\cite{shimizu} showed using $^1$H
nuclear magnetic resonance that
$\kappa$-(BEDT-TTF)$_2$Cu$_2$(CN)$_3$
did not undergo magnetic ordering
and that the temperature dependence of the uniform magnetic susceptibility
could still be fit by that for the triangular lattice.
However, it should be stressed that for these molecular
crystals the underlying triangular lattice of molecular dimers
(to which each spin is associated)
is not isotropic,\cite{mckenzie} and so it is important to know the
extent of the spatial anisotropy because this has a significant
effect on the possible ground state.
The isotropic triangular lattice is believed to be ordered,
but for $J_1/J_2$ = 0.7-0.9 the anisotropic
lattice could be quantum disordered.\cite{weihong}
Hence, determination of the actual ratio is
important for understanding these materials.


Here, we use high temperature series expansions to calculate the
temperature dependent uniform susceptibility of the spatially
anisotropic triangular-lattice models. Such calculations have been
done previously for the pure square and triangular-lattice cases
\cite{elstner,oitmaa} but not for the spatially anisotropic
triangular-lattice model. This method is particularly useful here,
as it allows one to cover the full range of $J_1/J_2$ ratios at
once. Our main finding is that the susceptibility, for these
antiferromagnets, shows a broad maximum at a temperature (which we
call the peak temperature $T_p$) of order the Curie-Weiss
temperature.  If the exchange constants are scaled to give the
same peak location, the magnitude of the peak susceptibility
varies with frustration. The unfrustrated models, represented by
the square-lattice and the linear-chain limits have similar peak
susceptibilities. The triangular-lattice deviates the most from
them, having a much smaller peak value, and a much flatter
temperature dependence. The parameter regimes, where the ground
states could be spin-disordered do not stand out in these calculations
\cite{weihong} and are similar to the triangular-lattice limit.
The reason for this is probably that at the temperature scales
considered the susceptibility is largely determined by
short-range frustration, rather than long length scale
physics such as the existence of spin liquid states at
zero temperature.

Comparison with the measured susceptibility of Cs$_2$CuCl$_4$ and
Cs$_2$CuBr$_4$ leads to exchange parameters in agreement with
previous neutron measurements. For the organic materials, it shows
that they are all close to the isotropic triangular-lattice limit.
But, some of them could be weakly anisotropic, leading to a quantum-disordered
ground state.
Since, the organic materials are close to a Mott metal-insulator
transition, we consider the possible role
of  multiple-spin exchange. Such interactions
can be necessary for a quantitative
description of such materials.\cite{coldea5}

\section{The frustrated model}
\begin{figure}[!htb]
\begin{center}
  \includegraphics[width=6cm,bbllx=0,bblly=0,bburx=232,
  bbury=153,angle=0,clip=]{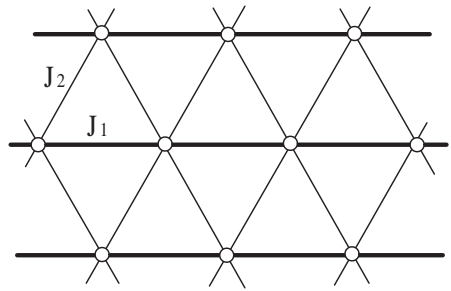}
  \caption{\label{fig1}The spatially  anisotropic
exchange constants for the Heisenberg model on the  triangular
lattice. The model can also be viewed as a square lattice with an
extra exchange along one diagonal.}
\end{center}
\end{figure}

The spatially anisotropic triangular-lattice is shown in Fig.\
\ref{fig1}. The antiferromagnetic Heisenberg model is described by
the Hamiltonian
\begin{equation}
H = J_1 \sum_{{\rm a}} {\bf S}_i\cdot {\bf S}_j  +
 J_2 \sum_{{\rm b}} {\bf S}_i\cdot {\bf S}_j \label{H}
\end{equation}
where the first sum runs over all nearest neighbor pairs along the
x-axis and the second sum runs over all other nearest neighbor pairs.
The vectors $\bf S$ represent spin-half operators.
It is evident that, for $J_1=0$, the model is equivalent to
the square-lattice Heisenberg model, for $J_2=J_1$ it is equivalent
to the isotropic triangular lattice model, and in the limit $J_2\to 0$,
it is equivalent to a model of decoupled linear chains.


We now discuss how we might quantify
how the amount of frustration in
the model varies with $J_2/J_1$.
Possible measures of frustration which have been discussed
before include:

1. The number of degenerate ground states.

2. How the competing interactions prevent the
  pairwise collinear alignment of spins that
  would give neighboring spins the lowest interaction energy.

In order to quantify 2., Lacorre\cite{lacorre}  considered
classical spins and introduced a ``constraint'' function
$F_c=-E_0/E_b$ which is the ratio of the ground state energy $E_0$
of the system to the ``base energy'', $E_b$ which is the sum of
all bond energies if they are independently fully satisfied,
i.e.

\begin{equation}
E_b = -\sum_{ij} |J_{ij}| (\bm{S}_i \cdot \bm{S}_j)_{\rm max}
\end{equation}

Lacorre suggested that $F_c$ has values ranging from -1 (no
frustration) to +1 (complete frustration). However, for spin
models that have a traceless Hamiltonian  the ground state energy
cannot become positive. So, $F_c$ must lie between -1
(unfrustrated) and 0 (fully frustrated)- the largest possible
value of $F_c$ is zero.
Considering a single isosceles triangular plaquette taken from the
lattice in Fig.\ \ref{fig1}, Lacorre found that for classical
(large-$S$) spins as a function of $J_2/J_1$, $F_c$ had its
maximum value ($-1/2$) for the isotropic triangle ($J_1=J_2$). The
same result holds for the infinite lattice.

Kahn\cite{kahn} recently stressed that for Heisenberg spins the
degeneracy of the ground state depends on the value of the spin
quantum number, $S$,  as well as the geometry of the plaquette.
For example, on an isotropic triangle, the ground state is
four-fold degenerate for $S=1/2$ but non-degenerate for
$S=1$.\cite{dai}
%
On a single isosceles triangle,
for $S=1/2$, the ground state has total spin $S_T=1/2$ and is
2-fold degenerate for  $J_1 \neq J_2$ and 4-fold degenerate at the
isotropic point $J_1=J_2$. We find that both $F_c$ is maximal
($-1/3$) and the ground state has the highest degeneracy for
$J_1=J_2$.
On the other hand, for spin $S=1$ the ground state is a
non-degenerate singlet ($S_T=0$) for a wide region near the
isotropic limit ($0.5<J_1/J_2<2$), is three-fold degenerate
($S_T=1$) outside this range ($J_1/J_2<0.5$ or $J_1/J_2>2$) and
has accidental 4-fold degeneracy at the special points
$J_1/J_2=0.5,2$. The function $F_c$ has no singular maximum, but a
plateau at $-0.5$ for the whole range $0.5<J_1/J_2<2$, so the
spin-1 case is much less frustrated than the extreme spin-$1/2$ case.

The above properties of the degeneracy and constraint
function are not unique to quantum spins but also hold for
the Ising model on the same
lattice. For a single isosceles triangle and for $S=1/2$ the
ground state energy changes at $J_1=J_2$ from $-J_1/4$ for
$J_1>J_2$, to $(-2J_2+J_1)/4$ for $J_1<J_2$. The ``base energy''
is $E_b = -(J_1 + 2 J_2)/4$ and hence $F_c$ has its maximum value
(-1/3) when $J_1=J_2$. The degeneracy of the ground state is 2
(only up-down symmetry) for $J_1<J_2$, 6 (only all up and all down
are not ground states) for $J_1=J_2$, and 4 (either one of the
$J_2$ bonds can be dissatisfied) for $J_1>J_2$. So indeed by both
measures for $J_1=J_2$ the model on a triangle is most frustrated.
Extending this analysis for a single triangle to a large lattice
of $N$ sites the difference is even more dramatic as the
degeneracy is\cite{wannier} $\exp(cN)$ for $J_1=J_2$, and is easily seen to be
only 2 for $J_1<J_2$ and $\exp(c^{\prime}N^{1/2})$ for $J_1>J_2$,
where $c$, $c^{\prime}$ are numbers of order one. So the model has
the largest ground-state degeneracy at the isotropic point.



Although, this paper is concerned with the quantum spin
model, the reason we mention the above properties of
classical models is because an important question is
whether our results concerning the connection
between the amount of frustration and the temperature
dependence of the susceptibility are also
exhibited by the corresponding classical Heisenberg and Ising models.
This may be the case if the temperature
dependence of the susceptibility
down to the peak is largely determined by the
frustration and correlations associated with a single placquette.

With regard to measures of frustration
we also note that from an experimental point of view
two measures that have been proposed previously.\cite{ramirez}
(i) The ratio of the Curie-Weiss temperature to the magnetic
ordering temperature. This increases with increasing frustration.
(ii) The amount of entropy at temperature scales much less
 than the exchange energy.

\section{High Temperature Series Expansions}

The high temperature series expansion method has been extensively
applied to and tested for quantum lattice models.\cite{elstner2}
We have obtained  high temperature expansions for arbitrary ratio
of $J_1/J_2$ to order $\beta^{10}$. We express the uniform
susceptibility, per mole, as
\begin{equation}
\chi= {N_A g^2\mu_B^2\over k T} \overline{\chi}
\end{equation}
where $N_A$ is Avogadro's number, $g$ the $g$-factor, $\mu_B$ a
Bohr-magneton, $k$ the Boltzmann constant and $T$ the absolute
temperature. The dimensionless quantity $\overline{\chi}$ can be
expressed in a high-temperature expansion in $J_2/T$ and
$y=J_1/J_2$, as \be \overline{\chi} =  \sum_{n=0} (J_2/T)^n
\sum_{m=0}^{n} c_{m,n} y^m/(4^{n+1}n!) \label{eq_chi} \ee The
integer coefficients $c_{m,n}$ complete to order $n=10$ are
presented in Table\ \ref{tabht}.


\section{Curie-Weiss Behavior and Beyond: Series Extrapolations}
\begin{figure}[!htb]
\begin{center}
  \includegraphics[width=6cm,bbllx=78,bblly=6,bburx=587,
  bbury=714,angle=-90,clip=]{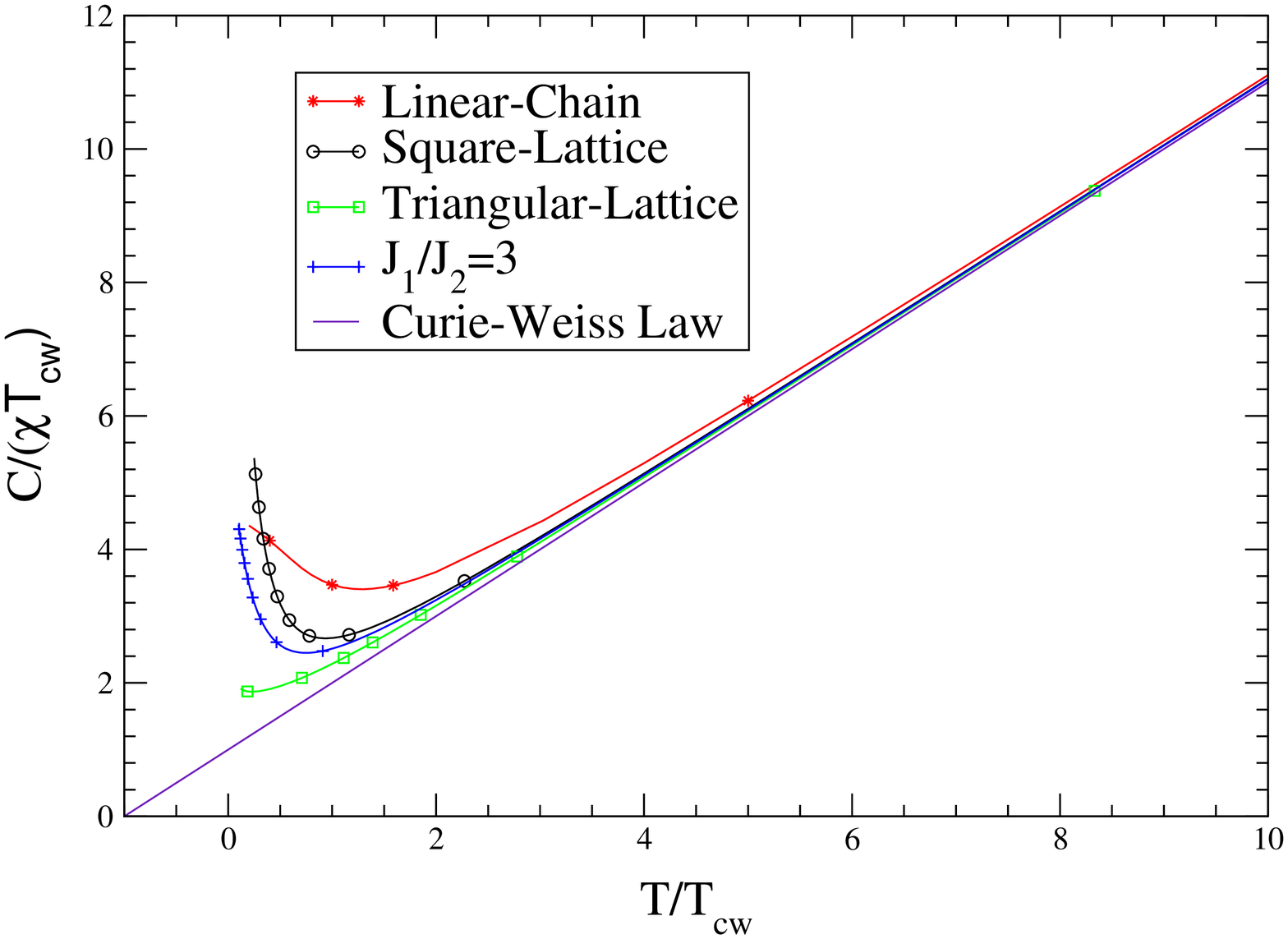}
  \includegraphics[width=6cm,bbllx=78,bblly=6,bburx=587,
  bbury=714,angle=-90,clip=]{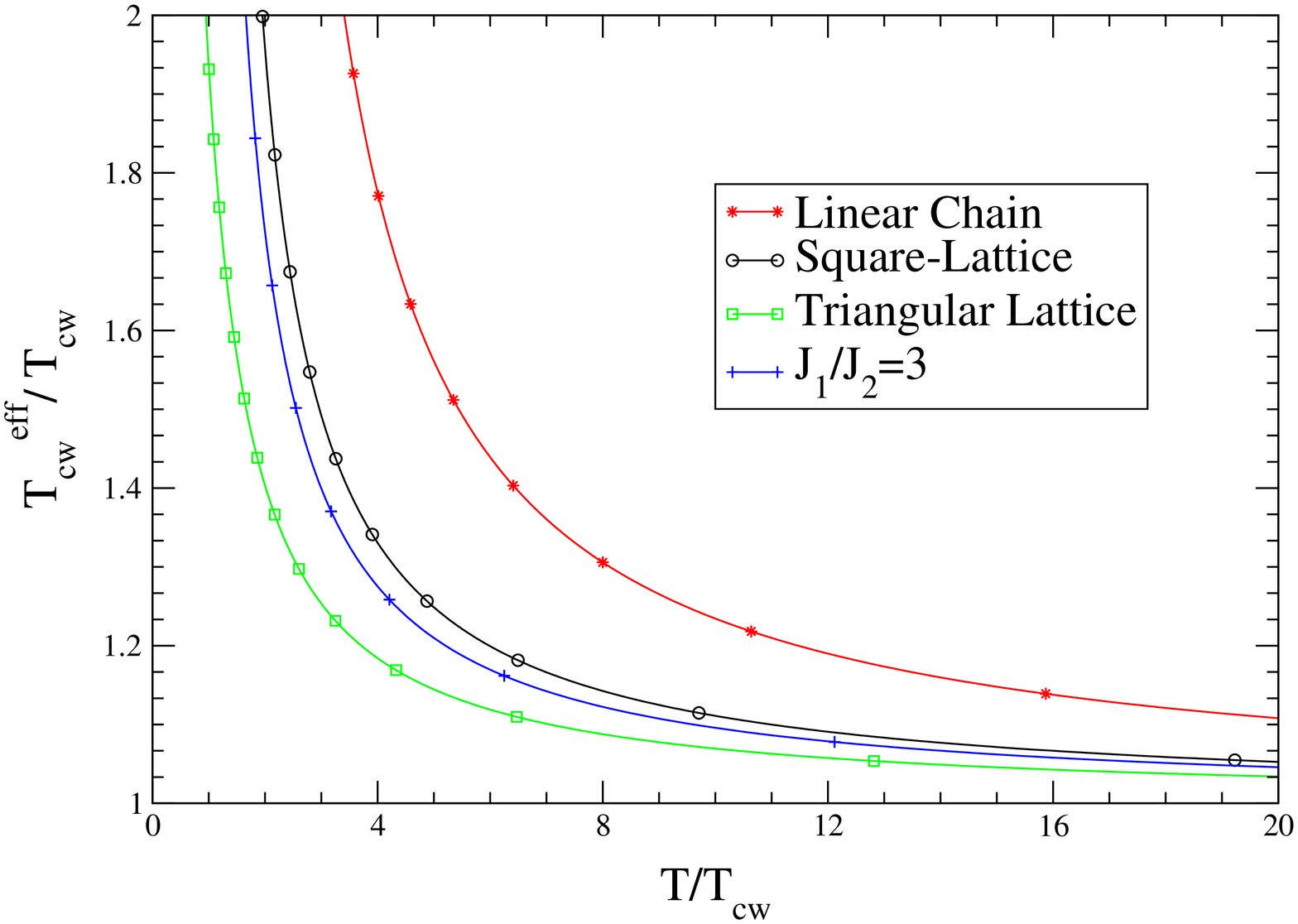}
  \caption{\label{fig2}
(Color online)(a) Inverse magnetic susceptibility as a function of
temperature in relative units of the Curie-Weiss constant $T_{\rm
cw}$. The susceptibility departs from the high-temperature
Curie-Weiss limit (Eq.(4)) already at temperatures a few times
$T_{\rm cw}$ due to short-range correlations. The smallest
departure occurs for the triangular lattice. (b) Effective
Curie-Weiss constant $T^{\rm eff}_{\rm cw}$ vs temperature found
by a local fit of the susceptibility to the Curie-Weiss form,
Eq.(7). The plot shows that fitting data below $4T_{\rm cw}$ can
result in large overestimates  of the Curie-Weiss constant.
}
\end{center}
\end{figure}

As is well known, the high temperature behavior of the susceptibility,
per mole, is given by a Curie-Weiss law
\begin{equation}
\chi={C\over T+T_{\rm cw}}
\end{equation}
For our model, the Curie constant
\begin{equation}
C=N_Ag^2\mu_B^2/4k=Ag^2,
\end{equation}
with $A=0.0938$ in cgs units. The Curie-Weiss temperature is
\begin{equation}
T_{\rm cw}=J_2+J_1/2.
\end{equation}
From an experimental point of view, an important question is: How
low in temperature is the Curie-Weiss law valid? To investigate
this, we plot in Fig.\ \ref{fig2}a, the normalized inverse
susceptibility as a function of $T/T_{\rm cw}$ for several
parameters, together with the Curie-Weiss law. It is clear that
below $T<10T_{\rm cw}$, the Curie-Weiss fit is no longer accurate.
Deviations from the Curie-Weiss behavior are the smallest near the
triangular-lattice limit, and largest for linear chains. If one
were to fit the inverse susceptibility below some temperature to a
Curie-Weiss behavior, one would get a systematically larger
Curie-Weiss temperature. To quantify this, we define an effective
temperature dependent Curie-Weiss constant $T_{\rm cw}^{\rm eff}$
as
\begin{equation}
T_{\rm cw}^{\rm eff} =-T - {\chi\over d\chi/dT}
\end{equation}
If one was to fit $\chi^{-1}$ to a linear curve in the vicinity of
some temperature ($T$) and use the intercept to estimate the
Curie-Weiss constant, one would get $T_{\rm cw}^{\rm eff}$. Fig.\
\ref{fig2}b shows how $T_{\rm cw}^{\rm eff}$ varies with
temperature for several parameter ratios. It shows that attempts
to fit to a Curie-Weiss behavior below four times the Curie-Weiss
temperature can result in an overestimate in the Curie-Weiss
constant by less than 20 percent for the isotropic triangular lattice,
whereas for the square lattice an error of 40 percent is possible.
Similar observations were made previously for the
classical Heisenberg antiferromagnet on a kagome lattice.\cite{harris}

To obtain the susceptibility for $T\le T_{cw}$, we need to develop
a series extrapolation scheme.
 We have used d-log Pad\'e and the integral differential
approximants to extrapolate the series \cite{gut,baker,fisher}. For the
linear chain model we use very long series given by Takahashi \cite{takahashi}
and for the square and triangular-lattice cases we have also used longer series.\cite{oitmaa,elstner}
In the former case, the calculated susceptibility agrees well with the exact
results obtained from Thermodynamic Bethe Ansatz calculations \cite{eggert}.
For the square and triangular-lattice cases it
also agrees well with previous numerical
calculations \cite{troyer,elstner}. In all cases, several integral/dlog-Pade
approximants are calculated, and in the plots below two outer approximants
are shown, {\it i.e.} a large number of approximants lie between
those shown. Based on our general experience with series extrapolations
\cite{example},
we feel confident that as long as the upper and lower curves are
not too far from each other, they bracket the true value of the thermodynamic
susceptibility. In general, we find that the extrapolations work well
down to the peak temperature and begin to deviate from each other below
the peak. It is not possible to address the zero and very
low-temperature behavior of the susceptibility from these calculations.

\begin{figure}[!htb]
\begin{center}
  \includegraphics[width=6cm,bbllx=78,bblly=6,bburx=587,
  bbury=714,angle=-90,clip=]{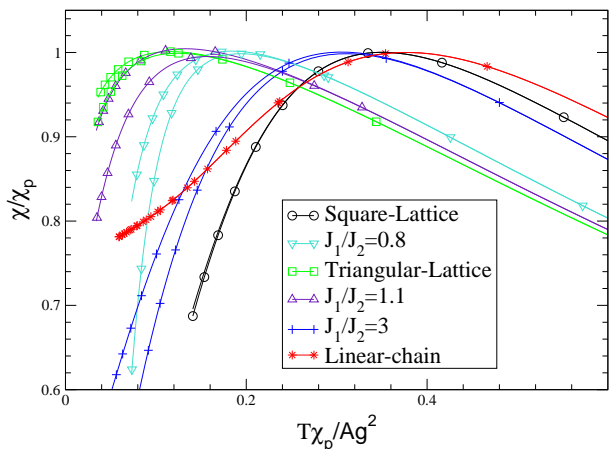} 
  \caption{\label{fig3}
(Color online) Susceptibility versus temperature for different
values of $J_1/J_2$. The peak susceptibility $\chi_p$ and the
Curie-Weiss constant $C=Ag^2$ are used to
define a dimensionless relative temperature scale. As discussed in
the text, the
two curves shown for each $J_1/J_2$ value are due to different
extrapolation schemes. For the most
frustrated triangular-lattice the peak in the susceptibility
occurs at the lowest relative temperature.}
\end{center}
\end{figure}

In Fig.\ \ref{fig3}, we show the uniform susceptibility, for
different $y=J_1/J_2$, as a function of temperature. For all
$J_1/J_2$ ratios, there are two plots showing the upper and lower
limits of extrapolated values as discussed in the previous
paragraph. The susceptibility is scaled to have a peak value of
unity, and the temperature axis is scaled by the peak
susceptibility to a dimensionless relative temperature. One finds
that the susceptibility peaks at a comparable relative temperature
for the unfrustrated square-lattice and linear chains. The primary
difference between these two models lies in the behavior of the
susceptibility below the peak. It decreases much more slowly for
the linear chains than it does for the square-lattice. We believe,
this is related to the fact that longer-range antiferromagnetic
correlations grow much faster for the square-lattice than they do
for linear-chains. Thus the shift of the spectral weight away from
zero wavevector  occurs more gradually for linear chains. For the
triangular-lattice, the peak is shifted to much lower relative
temperatures. Note that the triangular-lattice has a peak at a
temperature even lower than for $J_1/J_2=0.8$, where $T=0$
calculations show an absence of long range order \cite{weihong}.

\begin{figure}[!htb]
\begin{center}
  \includegraphics[width=6.5cm,bbllx=38,bblly=139,bburx=547,
  bbury=688,angle=0,clip=]{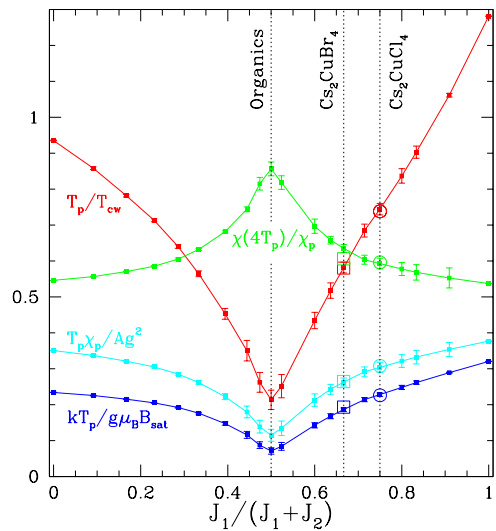} 
  \caption{\label{fig4}
(Color online) Variation of the key parameters of the
susceptibility curve $\chi(T)$ as a function of the frustration
ratio $J_1/(J_1+J_2)$: location of peak temperature $T_p$ relative
to the overall energy scale of the couplings, given by the
Curie-Weiss constant $T_{\rm cw}$ or the saturation field $B_{\rm
sat}$ required to overcome all antiferromagnetic interactions (see
text for more details), dimensionless product of peak
susceptibility and peak temperature $T_p \chi_p/A g^2 $ (with
$A=.0938$ in cgs units), and flatness of the susceptibility curve
$\chi(4T_p)/\chi(T_p)$. The isotropic triangular lattice
($J_1=J_2$) is the most frustrated with the lowest relative peak
temperature $T_p/T_{\rm cw}$, lowest peak susceptibility and
flattest curve at temperatures above the peak. The circles are
values extracted from the experimental
data for Cs$_2$CuCl$_4$ from Ref.\
\onlinecite{tokiwa} and the squares are for Cs$_2$CuBr$_4$ Ref.\
\onlinecite{ono}. This    suggests that the ratio $J_1/J_2$ is
close to $3.0$ and $2.0$, respectively, for these two materials.
The former is consistent with independent estimates from
neutron scattering.\cite{coldea4}
}
\end{center}
\end{figure}

From Fig.\ \ref{fig3} it is clear that frustration leads to a
reduction in the magnitude of the product $\chi_p T_p$ as well as
a reduction in the peak temperature $T_p$ with respect the
Curie-Weiss temperature $T_{\rm cw}$. These parameters are plotted
in appropriate dimensionless units in Fig.\ \ref{fig4} as a
function of the frustration ratio $J_1/(J_1+J_2)$, and both have a
minimum around the triangular lattice limit $J_1=J_2$. To connect
with experiments we also show the ratio of the peak temperature
$T_p$ and the Zeeman energy required to fully polarize the spins
$g\mu_BB_{\rm sat}$, related to the couplings strength
by\cite{coldea4}
\begin{equation}
g\mu_BB_{\rm sat}=
\left\{
\mbox{
\begin{tabular}{lcl}
$2J_1+2J_2+\frac{J^2_2}{2J_1}$ & for & $J_2 \leq 2 J_1$ \\
$4 J_2$ & for & $J_2 \geq 2J_1$~.
\end{tabular}
} \right.
\end{equation}
Fig.\ \ref{fig4} also shows the ratio $\chi(4T_p)/\chi_p$, which
is a measure of the flatness of the curves on the high-temperature
side of the peak. A larger value of this ratio implies a slower
decay of the susceptibility with temperature. These quantities
clearly show that the triangular-lattice is the most frustrated,
with the lowest peak temperature relative to the scale of the exchange
interactions, $T_p/T_{\rm cw}$ or $kT_p/g\mu_BB_{\rm sat}$, the
smallest dimensionless ratio $T_p\chi_p$ and the flattest peak
denoted by the largest $\chi(4T_p)/\chi_p$. The plots look very
symmetrical around the triangular-lattice limit, and there is
nothing anomalous about the case of $J_1/J_2=0.8$, where
zero-temperature studies give a disordered and gapped dimerized
ground state \cite{weihong}. We note that all of the extracted
parameters in Fig.\ \ref{fig4} are from the susceptibility curve
at temperatures above the peak and in order to see evidence for
the presence of a gap for $J_1/J_2\sim0.8$ as opposed to no gap in
the isotropic triangular-lattice case one would be required to
analyze the susceptibility curve at temperatures much below the
estimated gap $\Delta\sim~0.25J_2\sim 0.5~T_p$ in
the dimerised state,\cite{weihong} and such low
temperatures are not accessible by the present series
calculations.


\section{Comparison with Experimental Systems}

\begin{figure}[!htb]
\begin{center}
  \includegraphics[width=6cm,bbllx=78,bblly=6,bburx=587,
  bbury=714,angle=-90,clip=]{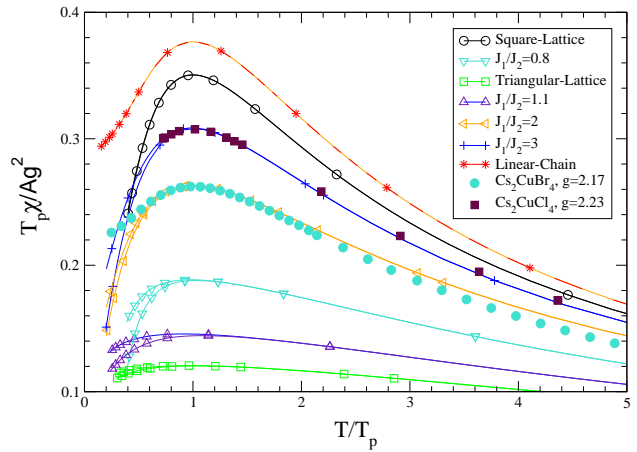}
  \caption{\label{fig5}
(Color online) Susceptibility vs temperature in units of the peak
temperature $T_p$. The isotropic triangular lattice (green line)
has the lowest and flattest susceptibility. Solid squares show
data points for the anisotropic triangular lattice material
Cs$_2$CuCl$_4$ ($a$-axis, Ref.\ \onlinecite{tokiwa}), solid
circles show data for Cs$_2$CuBr$_4$ from Ref.\ \onlinecite{ono}.
}
\end{center}
\end{figure}

In this section, we compare our theoretical results with
experimental data on Cs$_2$CuCl$_4$, Cs$_2$CuBr$_4$ and various
organic materials. In Fig.\ \ref{fig5}, we show the susceptibility
as a function of temperature for different $J_1/J_2$ ratio, where
the temperature is scaled by the peak temperature ($T_p$) and the
susceptibility itself is scaled by the peak temperature to give a
dimensionless reduced susceptibility. This plot is very
instructive as it allows one to clearly read out the $J_1/J_2$
ratios. Also shown are the susceptibilities for the materials
Cs$_2$CuCl$_4$ and Cs$_2$CuBr$_4$, with their g-values taken from
ESR experiments \cite{esr1,esr2}. In this plot with no free
parameters, it is apparent that the $J_1/J_2$ ratio is near $3.0$
for Cs$_2$CuCl$_4$ and near $2.0$ for Cs$_2$CuBr$_4$. Some of
these results can also be seen from Fig.\ \ref{fig4}, where key
dimensionless ratios of the temperature dependent susceptibility
are shown.

\begin{figure}[!htb]
\begin{center}
  \includegraphics[width=6cm,bbllx=78,bblly=6,bburx=587,
  bbury=714,angle=-90,clip=]{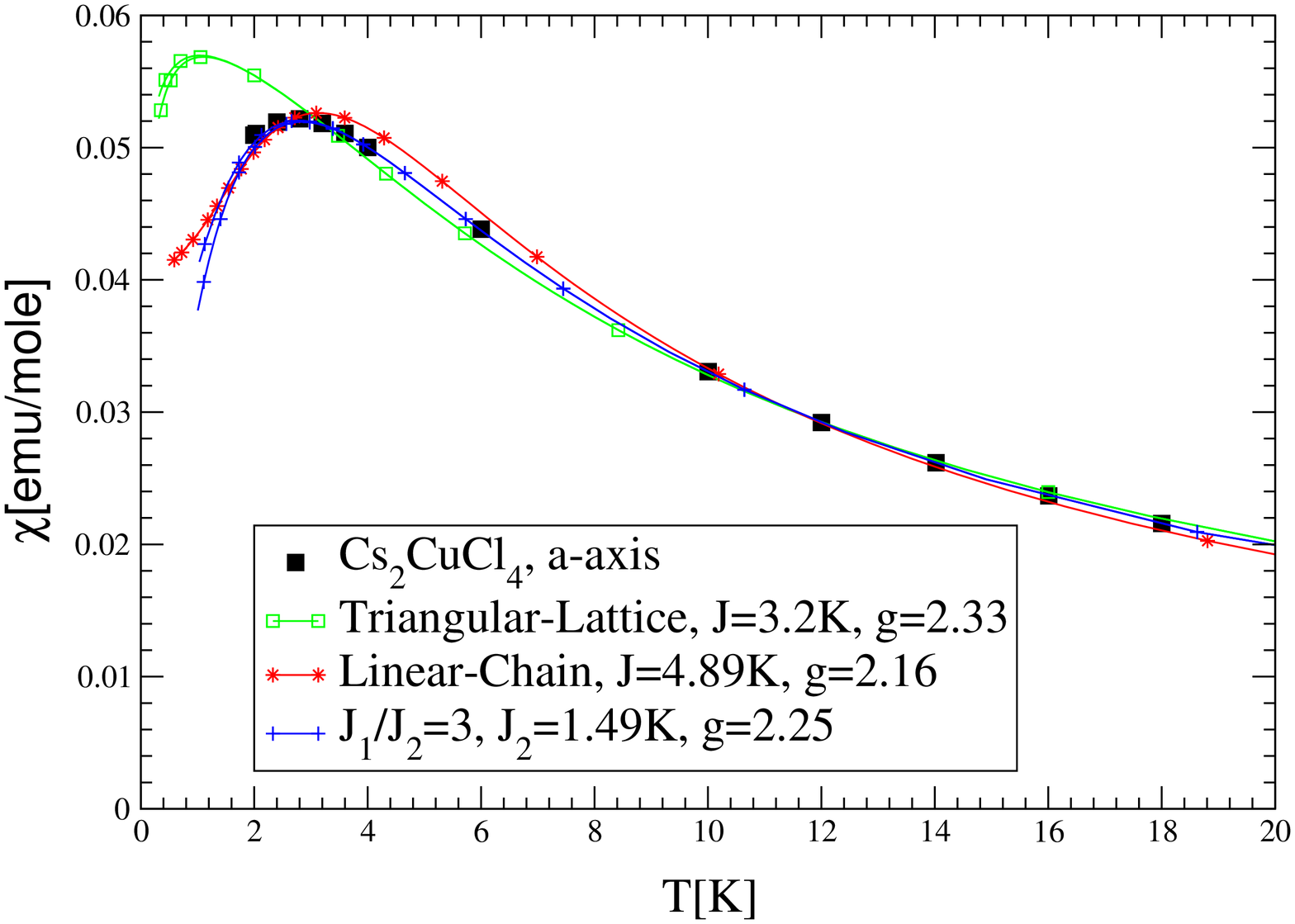}
  \includegraphics[width=6cm,bbllx=78,bblly=6,bburx=587,
  bbury=714,angle=-90,clip=]{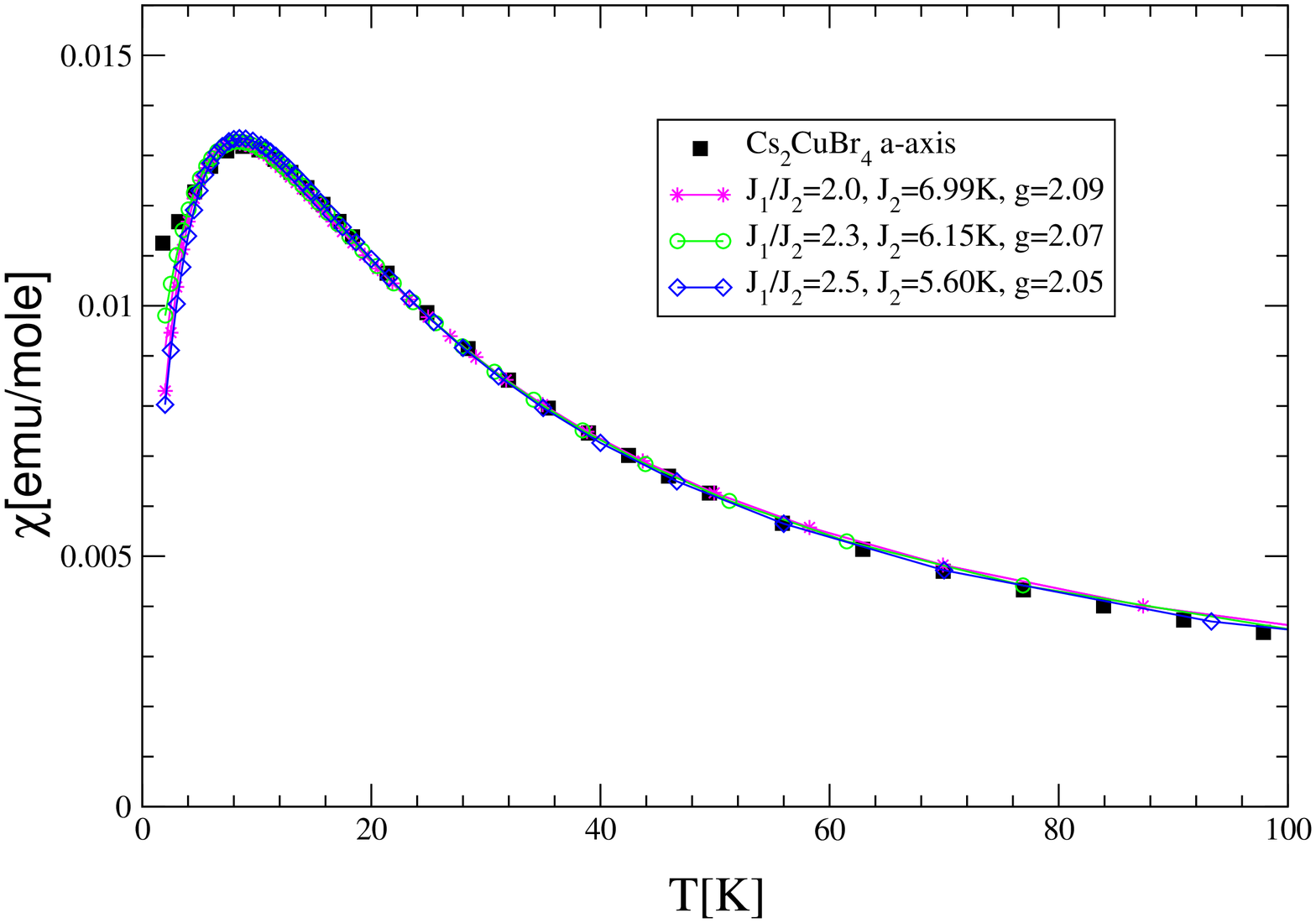}
  \caption{\label{fig6}
(Color online) Fits of the susceptibility in Cs$_2$CuCl$_4$ (a)
and Cs$_2$CuBr$_4$ (b), see text. }
\end{center}
\end{figure}

A more detailed comparison of the susceptibility for the
materials, Cs$_2$CuCl$_4$ and Cs$_2$CuBr$_4$, allowing $g$ to vary
freely is shown in Fig.\ \ref{fig6}. Once $g$ is allowed to vary,
the material Cs$_2$CuCl$_4$ can be fit above the peak not too
badly even with the pure square-lattice model (not shown). However, a much
improved fit happens with $J_1/J_2=3$ and $J_2=1.49$~K in
excellent agreement with the exchange values extracted directly
from neutron scattering measurements.\cite{coldea4}
Also shown are fits to linear chain and triangular-lattice limits,
which bracket $J_1/J_2=3$. One can see significant deviation in both
limits. The large deviation from the isotropic triangular-lattice case
shows that frustration is relatively weak in this material.

For Cs$_2$CuBr$_4$, the best fit for $J_1/J_2=2$ arises with
$J_2=6.99$~K. However, when $g$ is allowed to vary,
a range of $J_1/J_2$ values from $1.8$
to $2.8$ give comparable fits, several of which are shown in figure.
In general, the high temperature data is better fit by a larger
$J_1/J_2$ value, whereas the data at and around the peak is better
fit by a smaller $J_1/J_2$ value. No choice of parameters can fit the
very low temperature data (below half the peak temperature).
These values are also consistent with previous
estimates. Using the value of the incommensurate ordering
wavevector $\bm{Q}=0.575(1)\bm{b}^*$ observed by neutron
scattering,\cite{ono} classical spin-wave theory gives
$J_1/J_2=2.14$ whereas including quantum renormalization
corrections as predicted by large-$N$ Sp($N$) theory\cite{chung}
gives $J_1/J_2\sim1.8$, and series expansions\cite{weihong} gives
$J_1/J_2\sim1.4$. This calls into question the rather large
renormalization of the ordering wavevector found in the series
expansion study.

\begin{figure}[!htb]
\begin{center}
  \includegraphics[width=6cm,bbllx=78,bblly=6,bburx=587,
  bbury=714,angle=-90,clip=]{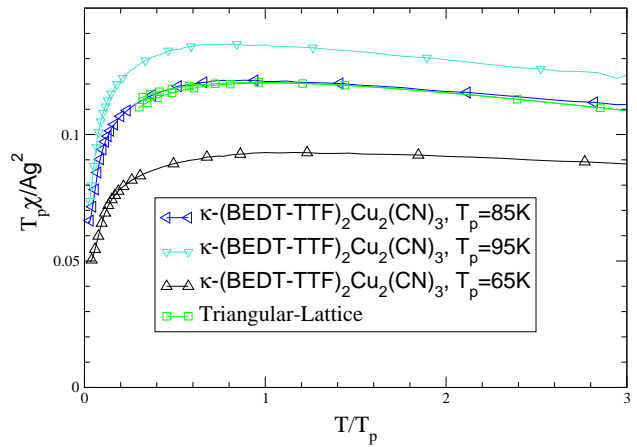}
  \caption{\label{fig7}
(Color online) Comparison of the temperature dependence of the
magnetic susceptibility of  $\kappa$-(BEDT-TTF)$_2$Cu$_2$(CN)$_3$
with series expansions calculations for isotropic
triangular-lattice. The experimental data is from Ref.\
\onlinecite{shimizu}. A value of g=2.006 was used based on
electronic spin resonance measurements.\cite{komatsu} We see that
this material is well described by a Heisenberg model on the
isotropic triangular lattice, with peak temperature $T_p=85$~K.
 Note also that that the
agreement is quite sensitive to changes in the value of $T_p$, a
quantity that is difficult to pin-point in a flat curve.}
\end{center}
\end{figure}

Now we turn to the organic materials. In Fig.\ \ref{fig7}, we
show a corresponding comparison for the material
$\kappa$-(BEDT-TTF)$_2$Cu$_2$(CN)$_3$. Only the theoretical data
for the isotropic triangular-lattice are shown. One can see an
important difficulty in using the $T_p$-scaled plots near the
triangular-lattice limit to determine $J_1/J_2$. For the organic
material, $\kappa$-(BEDT-TTF)$_2$Cu$_2$(CN)$_3$, the measured
susceptibility is very flat and it is difficult to determine the
peak temperature $T_p$. From the data, the peak temperature
appears to be between $65$~K and $95$~K. Using the values for
$T_p$ of $65$~K and $95$~K, one can either get the data to fall
above or below the triangular-lattice values. A suitably chosen
peak temperature allows one to get very close agreement with the
triangular-lattice limit. This peak temperature can also be used
to determine the exchange constant. However, for the
triangular-lattice, there is theoretical uncertainty in the peak
location. Hence, it is more accurate to directly fit the
experimental data to theory to obtain the exchange constants. For
$\kappa$-(BEDT-TTF)$_2$Cu$_2$(CN)$_3$, fixing $g=2.006$ and
$J_1=J_2$, the best fit leads to $J_1=256$~K, a value close to
that obtained by Shimizu {\it et al}.\cite{shimizu}

\begin{figure}[!htb]
\begin{center}
 \includegraphics[width=6cm,bbllx=78,bblly=6,bburx=587,
  bbury=714,angle=-90,clip=]{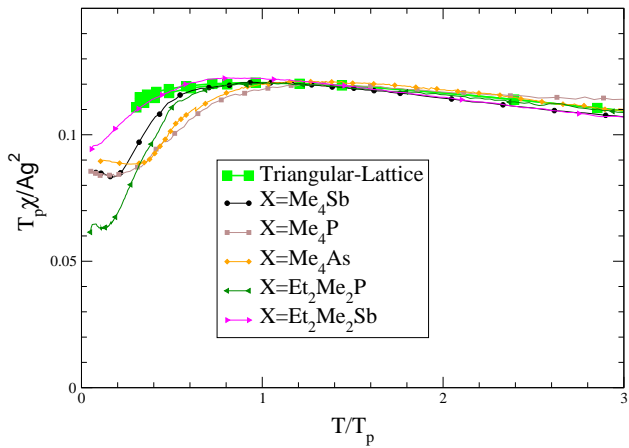}
 \caption{\label{fig8}
(Color online) Comparison of the temperature dependence of the
magnetic susceptibility of five different organic molecular
crystals from the family $\beta'$-[Pd(dmit)$_2$]X (different X are
indicated with Me=CH$_3$, Et=C$_2$H$_5$) with series expansions
for isotropic triangular-lattice. Experimental data is from Ref.\
\onlinecite{tamura}. A value of $g=2.04$ was used based on
electronic spin resonance measurements.\cite{nakamura} All of these
materials are well described by a Heisenberg model close to that
for the isotropic triangular lattice, assuming that ring-exchange
interactions do not need to be taken into account.
}
\end{center}
\end{figure}

The ability to fit flat susceptibilities to the isotropic
triangular-lattice model is further illustrated in Fig.\
\ref{fig8}, where the susceptibility data are shown from five
different molecular crystals in the family
 $\beta'$-[Pd(dmit)$_2$]X
(where dmit is the electron acceptor molecule
thiol-2-thione-4, 5-dithiolate, C$_3$S$_5$)
and the cation
X = Me$_4$As, Me$_4$P, Me$_4$Sb, Et$_2$Me$_2$P, and Et$_2$Me$_2$Sb,
 where Me = CH$_3$ and Et = C$_2$H$_5$, denote methyl and
 ethyl groups, respectively). We have taken the $g$-value to
be $2.04$. By adjusting the peak temperature, they can all be
brought to rough agreement with the triangular-lattice model. Assuming
isotropic interactions, and $g=2.04$, we estimate the exchange
constants to be 283~K, 289~K, 270~K, 279~K and 247~K respectively.
It is clear that none of these organic materials are far from the
isotropic triangular-lattice limit. But, we emphasize, that by
this method it is difficult to discriminate between $J_1/J_2$
ratios in the range $.85<J_1/J_2<1.15$. Note that the latter
regions also include quantum disordered phases.

\begin{figure}[!htb]
\begin{center}
  \includegraphics[width=6cm,bbllx=78,bblly=6,bburx=587,
  bbury=714,angle=-90,clip=]{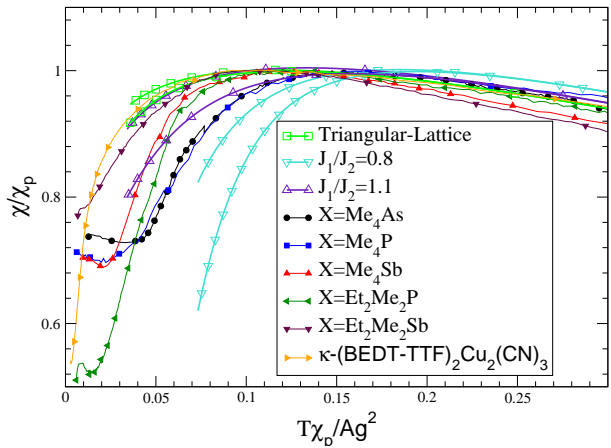}
  \caption{\label{fig9}
(Color online) Parameter free comparison of the susceptibility
data on organic materials with theoretical plots scaled by the
peak susceptibility, $\chi_p$, which is easy to measure accurately
for a flat curve. It is evident that the materials deviate only
slightly from the isotropic triangular-lattice model and have
$J_1/J_2$ ratios in the range $0.85$ to $1.15$. }
\end{center}
\end{figure}

To avoid the problem of determining the peak temperature, we go
back to Fig.\ \ref{fig3}, and scale the data by the peak
susceptibility. These can be inferred accurately from the data,
even when the peak temperature cannot. In Fig.\ \ref{fig9}, we
show such a comparison of experimental data with theory. The data
for $\kappa$-(BEDT-TTF)$_2$Cu$_2$(CN)$_3$ lies extremely close to
the isotropic triangular-lattice case. The other materials deviate
from the $J_1=J_2$ limit, but still lie in the range
$0.85<J_1/J_2<1.15$. If we assume that the systems are described
by the isotropic triangular-lattice, the exchange constant can be
read of from the peak susceptibility by using the relation
$J=0.0035 g^2/\chi_p$. This leads to exchange constants of
250K for $\kappa$-(BEDT-TTF)$_2$Cu$_2$(CN)$_3$ and 280~K, 289~K,
260~K, 273~K and 236~K for the other materials. These values are
close to those obtained from the best fits.

It should be noted here that in the experimental data, a
Curie term from magnetic impurities and a diamagnetic term
has been subtracted and these can also influence the
determination of exchange parameters. However, it is unlikely that
any of these materials are very far from the isotropic triangular
lattice limit.

From the fits the Heisenberg couplings are comparable for all
materials and around $250$~K. We now consider how these compare
with quantum chemistry calculations. The exchange constants can be
related to parameters in an underlying Hubbard
model\cite{mckenzie,tamura,kato2} where $J=2t^2/U$ and $t$ is the
intersite (i.e., inter-dimer) hopping and $U$ is the cost of
double occupancy for two electrons or holes on a dimer. If the
Coulomb repulsion $U_0$ on a single molecule within the dimer is
much larger than the inter-molecular hopping $t_0$ within a dimer
then $U \simeq 2 t_0$. For
 $\beta'$-[Pd(dmit)$_2$]X
electronic structure calculations based on the
local-density approximation (LDA)\cite{kato2,tamura,tohno} give
$t \sim 30$ meV and $t_0 \sim 500$ meV, and
so $J \sim 50$ K.
For $\kappa$-(BEDT-TTF)$_2$Cu$_2$(CN)$_3$
H\"uckel electronic structure calculations give
$t \sim 50$ meV and $t_0 \sim 200$ meV,\cite{komatsu,merino3}.
The resulting $U \simeq 400$ meV is comparable to that
deduced from measurements of the frequency dependent optical conductivity
of similar $\kappa$ materials.\cite{mckenzie,girlando}
This value of $U$ is smaller than values deduced
from quantum chemistry calculations on
isolated dimers, which do not take into account
screening.\cite{qcref}
 Using the above values of $t$ and $U$ gives $J \sim 100$ K.


Note that the quality of fit is best for the material
$\kappa$-(BEDT-TTF)$_2$Cu$_2$(CN)$_3$, where it really fits well
with the isotropic triangular-lattice model. However, it is
also a system that does not order down to very low temperatures.\cite{shimizu}
This remains a puzzle.
The quality of fits was not as good for the other
organic compounds. It is quite possible that the organics
have other interactions not captured by the Heisenberg model. In a
Mott insulator when a perturbation expansion in $t/U$ is used to
derive an effective Hamiltonian for the spin degrees of freedom
one finds that to four order in $t/U$ there are cyclic exchange
terms in the Hamiltonian.\cite{macdonald} If $U/t < 10$ then these
terms may be important. Recent neutron scattering studies showed
the effect of such interactions on the dispersion of spin
excitations in La$_2$CuO$_4$.\cite{coldea5}

The metallic
phase of the organics are
in the regime $U/t \sim 5-10$ [Ref. \onlinecite{mckenzie}] and so
one might expect multiple-exchange terms to be relevant
in the insulating phase. For the triangular
lattice triple exchange is also possible. However, for spin-1/2
this just corresponds to a renormalisation of the
nearest-neighbour two-particle exchange.\cite{misguich}
The frustrating effects of multiple-spin exchange on
the isotropic triangular lattice lead to rich physics and have an
experimental realisation in monolayers of solid $^3$He on
graphite.\cite{masutomi}
Let $J$ denote the nearest-neighbour exchange  and
$J_4$ the multiple
spin-exchange, involving the four spins comprising
a pair of triangular placquettes.
This model has been studied extensively and exact
diagonalisation calculations suggest that the 120 degree Neel
state, which is the ground state
for $J_4=0$, is destroyed when $J_4 > 0.1J$.\cite{ming}
It is appealing to think that this could be
the explanation for why
$\kappa$-(BEDT-TTF)$_2$Cu$_2$(CN)$_3$
does not magnetically order, whereas it should if it is
really described
by the isotropic triangular lattice nearest neighbour model.
This material is close to a Mott-Hubbard metal-insulator
transition since the insulating state is destroyed
under pressure or uniaxial stress.\cite{komatsu,sc-dmit}
However, it is not clear that $J_4$ will be large enough
in the actual material. The expressions
derived from a $t/U$ expansion give\cite{macdonald}
$J_4/J  = 10 (t/U)^2$.
This means one must have $U < 8t$ to obtain a spin liquid.
However, exact diagonalisation of the Hubbard model
on the isotropic triangular lattice at half filling, shows
that the insulating state only occurs for $U > 12 t.$\cite{capone}
Hence, it is not clear   that multiple-spin exchange
could account for the fact that this material appears
to be close to the isotropic triangle but does not magnetically order.
However, to definitively resolve this issue would require
a detailed study of the spatially  anisotropic model
with four-spin exchange.

\section{Conclusions}
In this paper, we have developed high temperature expansions for the
uniform susceptibility of the spatially anisotropic triangular lattice
Heisenberg model. We find that the temperature dependence of the
susceptibility at temperatures of order the exchange constants,
are sensitive to frustration, that is, the ability
of spins to align antiparallel to all their neighbors.
The square-lattice and linear chain limits have similar
reduced susceptibilities at and above the peak,
while the triangular-lattice limit appears most
frustrated, with the smallest and flattest susceptibilities.
Comparison with various experimental systems shows that
a variety of organic materials are close to the isotropic triangular-lattice
limit, whereas the inorganic materials Cs$_2$CuCl$_4$ and
Cs$_2$CuBr$_4$ are much less frustrated.

It would be nice to have a simple formalism which could provide an
analytic relation between the peak susceptibility and exchange
parameters. Qualitatively, our arguments show that
short-range frustration, or the inability to align parallel
with respect to neighbors as quantified by the parameter $F_c$
in Section II, is maximum near the isotropic limit and this
is what pushes the peak in the susceptibility down to
lower temperatures. For a wide
range of frustrated antiferromagnets it has been previously
pointed out that the Curie-Weiss law holds to relatively low
temperatures.\cite{ramirez,schiffer} Several  theoretical models,
mostly for classical spins, have been developed to explain
this\cite{moessner3,huber}. Basically, frustration leads to
individual plaquettes or spin clusters behaving essentially
independently. However, our models are less frustrated than that
and hence always develop substantial correlations. This means that
any simplistic explanation is unlikely.


In organic molecular crystals a weak temperature dependence
of the magnetic susceptibility is often interpreted as
being evidence for metallic behavior, since
for a Fermi liquid the susceptibility is weakly temperature
dependent. However, this is inconsistent with the fact
that in most of these materials above temperatures
of about 50 K there is no Drude peak
in the optical conductivity and the resistivity has
a non-monotonic temperature dependence and values of
order the Mott limit.\cite{mckenzie,merino2}
This work shows that due to the substantial magnetic frustration
the susceptibility can actually be due to local magnetic moments,
even though in the range up to 300 K one does not see a clear
Curie temperature dependence.

In a future study we will consider the temperature dependence of
the specific heat capacity for this model. A previous
study\cite{bernu2} of the square lattice, single chain, and
triangular lattice Heisenberg model found that the peak in the
specific heat versus temperature curve occurred around $J$ for all
models but was much broader for the triangular lattice. A related
issue was that as the temperature decreases the entropy decreases
much more slowly for the triangular lattice than the others.

\acknowledgements

We thank T. Ono, Y. Tokiwa, M. Tamura, and Y. Shimizu for sending
us their experimental data. We thank R. Moessner and B. Powell
for a critical reading of the manuscript. We have also benefitted from
discussions with S. Bramwell, B. Powell, and D. McMorrow.
 This work is supported by the
Australian Research Council (ZW and RHM), the US National Science
Foundation grant number DMR-0240918 (RRPS) and the United Kingdom
Engineering and Physical Sciences Research Council grant number
GR/R76714/01 (RC). RHM thanks UC Davis, ISIS, Rutherford
Appleton Laboratory,
 and the Clarendon Laboratory, Oxford University for hospitality.
 We are grateful for the computing resources provided
 by the Australian Partnership for Advanced Computing (APAC)
National Facility and by the
Australian Centre for Advanced Computing and Communications (AC3).


\begin{thebibliography}{99}

\bibitem{aeppli} G.
Aeppli and P. Chandra,
Science {\bf 275},  177 (1997).

\bibitem{senthil}
 T. Senthil, A. Vishwanath, L. Balents, S. Sachdev, and M. P. A. Fisher,
  Science {\bf 303}, 1490 (2004).

\bibitem{misguich2}
For a review of quantum antiferromagnets
in two dimensions, see
G. Misguich and C. Lhuillier, in
``Frustrated spin systems'', edited by H. T. Diep, World-Scientific (2003);
cond-mat/0310405.


\bibitem{greedan}
 J. E. Greedan, J. Mater. Chem. {\bf 11}, 37 (2001).

\bibitem{harrison}
 For a review of frustrated magnets
 from a synthesis point of view  see, A.
Harrison, J. Phys.: Condens. Matter {\bf 16}, S553 (2004).

\bibitem{ramirez}
A.P. Ramirez, Annu. Rev. Mater. Sci. {\bf 24}, 453 (1994).

\bibitem{rvb}
P. W. Anderson, Science {\bf 235}, 1196 (1987).

\bibitem{anderson} P. W. Anderson, Mater. Res. Bull. {\bf 8}, 153 (1973);
P. Fazekas and P. W. Anderson, Philos. Mag. {\bf 30}, 423 (1974).

\bibitem{wannier} G. H. Wannier, Phys. Rev. {\bf 79}, 357 (1950).


\bibitem{coldea2}
R. Coldea,
D. A. Tennant, A. M. Tsvelik,
 and Z. Tylczynski, Phys. Rev. Lett. \textbf{86}, 1335 (2001);
R. Coldea, D. A. Tennant, and Z. Tylczynski Phys. Rev. B
\textbf{68}, 134424 (2003).

\bibitem{ono}
T. Ono, H. Tanaka, H. Aruga Katori, F. Ishikawa, H. Mitamura, and
T. Goto, Phys. Rev. B \textbf{67}, 104431 (2003).

\bibitem{mckenzie}
R. H. McKenzie, Comments Condens. Matter Phys. \textbf{18}, 309
(1998).

\bibitem{natio}
S.J. Clarke, A.J. Fowkes, A. Harrison, R.M. Ibberson, and M.J.
Rosseinsky, Chem. Mater. {\bf 10}, 372 (1998).

\bibitem{suzuki}
M. Suzuki, I. S. Suzuki, C. R. Burr, D. G. Wiesler, N. Rosov, and K. Koga,
Phys. Rev. B \textbf{50}, 9188   (1994).

\bibitem{bramwell}
S. T. Bramwell, S. G. Carling, C. J. Harding, K. D. M. Harris, B. M. Kariuki, L. Nixon, and I. P. Parkin,
 J. Phys.: Condens. Matter {\bf 8}, L123 (1996).

\bibitem{laughlin} V. Kalmeyer and R. B. Laughlin, Phys. Rev. Lett.
{\bf 59}, 2095 (1987).

\bibitem{sondhi} R. Moessner and S. L. Sondhi, Phys. Rev. Lett. {\bf 86},
1881 (2001).

\bibitem{lhuillier} B. Bernu, P. Lecheminant, C. Lhuillier and  L. Pierre,
Phys. Rev. B {\bf 50}, 10048 (1994).

\bibitem{huse} R. R. P. Singh and D. A. Huse, Phys. Rev. Lett. {\bf 68},
1766 (1992).

\bibitem{farnell} D. J. J. Farnell, R. F. Bishop and K. A. Gernoth,
Phys. Rev. B \textbf{63}, 220402 (2001).

\bibitem{sorella}
L. Capriotti, A. E. Trumper, and S. Sorella
Phys. Rev. Lett. {\bf 82}, 3899 (1999).

\bibitem{weihong} W. Zheng, R. H. McKenzie and R. R. P. Singh,
Phys. Rev. B{\bf 59}, 14367 (1999).

\bibitem{row}
The classical version of this model was first studied in H.
Kawamura, Prog. Theor. Phys. Suppl. {\bf 101}, 545 (1990); W. Zhang, W.
M. Saslow and M. Gabay, Phys. Rev. B {\bf 44}, 5129 (1991).

\bibitem{stephenson}
The thermodynamic properties of
the Ising version of this model were found exactly by
R. M. F. Houtappel, Physica {\bf 16}, 425 (1950).
The exact  correlation functions were found and analyzed
extensively
by J. Stephenson, J. Math. Phys. {\bf 5}, 1009 (1964);
 ibid. {\bf 11}, 420 (1970).

\bibitem{merino}
A. E. Trumper, Phys. Rev. B {\bf 60}, 2987 (1999); J. Merino,
R. H. McKenzie, J. B. Marston, and  C.-H. Chung,
 J. Phys.: Condens. Matter
 \textbf{11}, 2965 (1999);
M.Y. Veillette, J.T. Chalker, and R. Coldea,
cond-mat/0501347.


\bibitem{chung}
C.-H. Chung, J. B. Marston, and R. H. McKenzie,
 J. Phys.: Condens. Matter
  \textbf{13}, 5159 (2001);
C.-H.  Chung, K.    Voelker, and Y.   B.   Kim, Phys. Rev. B
\textbf{68}, 094412 (2003).

\bibitem{wen}
 Y. Zhou and X.-G. Wen,
cond-mat/0210662.

\bibitem{ceccatto}
L. O. Manuel and H. A. Ceccatto,
 Phys. Rev. B \textbf{60}, 9489 (1999).

\bibitem{yunoki}
S. Yunoki and S. Sorella, Phys. Rev. Lett. \textbf{92}, 157003
(2004).

\bibitem{coldea4}
R. Coldea, D. A. Tennant,
 K. Habicht, P. Smeibidl, C. Wolters, and Z. Tylczynski
Phys. Rev. Lett. \textbf{88}, 137203 (2002).

\bibitem{emery}
G. Castilla, S. Chakravarty, and V. J. Emery, Phys. Rev. Lett.
\textbf{75}, 1823 (1995).

\bibitem{fabricius}
For a detailed discussion see,
K. Fabricius, A. Klumper, U. Low, B. Bochner, T. Lorenz,
 G. Dhalenne, and A. Revcolevschi,
Phys. Rev. B \textbf{57}, 1102 (1998); A. Buhler, U. Low, and G.
S. Uhrig Phys. Rev. B \textbf{64}, 024428 (2001).

\bibitem{powell}
J. Y. Gan, Yan Chen, Z. B. Su, and F. C. Zhang,
cond-mat/0409482;
B. J. Powell and R. H. McKenzie, cond-mat/0410125, to appear
in Phys. Rev. Lett.;
J.  Liu, J.   Schmalian, N, Trivedi,
cond-mat/0411044.


\bibitem{tamura}
M. Tamura and R. Kato, J. Phys.: Condens. Matter \textbf{14}, L729
(2002).

\bibitem{shimizu}
Y. Shimizu, K. Miyagawa, K. Kanoda, M. Maesato, and G. Saito,
Phys. Rev. Lett. \textbf{91}, 107001 (2003).

\bibitem{elstner} N. Elstner, R. R. P. Singh and A. P. Young,
Phys. Rev. Lett. \textbf{71}, 1629-1632 (1993); N. Elstner, R. R.
P. Singh and A. P. Young, J. Appl. Phys. {\bf 75}, 5943 (1994).

\bibitem{oitmaa}J. Oitmaa and E. Bornilla, \prb {\bf 53}, 14228 (1996).

\bibitem{coldea5}
R. Coldea, S.M. Hayden, G. Aeppli, T.G. Perring, C.D. Frost, T.E.
Mason, S.W. Cheong, and Z. Fisk, Phys. Rev. Lett. {\bf 86}, 5377
(2001).

\bibitem{lacorre}
 P. Lacorre, J. Phys. C {\bf 20}, L775 (1987).



\bibitem{kahn}
 O. Kahn, Chem. Phys. Lett. {\bf 265}, 109 (1997).

\bibitem{dai}
 D. Dai and M.-H. Whangbo, J. Chem. Phys. {\bf 121}, 672 (2004).

\bibitem{elstner2}
For a review see, N. Elstner, Int. J. Mod. Phys. B {\bf 11}, 1753
(1997).

\bibitem{harris}
A. B. Harris, C. Kallin, and A. J. Berlinsky,
 Phys. Rev. B \textbf{45}, 2899 (1992),
(see Fig. 14 and equation (4.5)).

\bibitem{gut}A.J. Guttmann, in ``Phase Transitions and Critical
Phenomena'', Vol. \textbf{13} ed. C. Domb and J. Lebowitz (New
York, Academic, 1989).

\bibitem{baker} D. L. Hunter and G. A. Baker, Jr., Phys. Rev.
B {\bf 19}, 3808 (1979).

\bibitem{fisher} M. E. Fisher and H. Au-Yang, J. Phys. A {\bf 12},
1677 (1979).

\bibitem{takahashi}
M. Shiroishi and M. Takahashi, \prl {\bf 89},
 117201 (2002).

\bibitem{eggert} S. Eggert, I. Affleck and M. Takahashi,
Phys. Rev. Lett. {\bf 73}, 332 (1994).

\bibitem{troyer} J. K. Kim amd M. Troyer, Phys. Rev. Lett. {\bf 80},
2705 (1998).

\bibitem{example} See for example N. Elstner {\it et. al.}, Phys. Rev.
Lett. {\bf 75}, 938 (1995).

\bibitem{tokiwa}
Y. Tokiwa {\it et al.} (in preparation).

\bibitem{esr1}E.J. Rzepniewski, PhD Thesis, Oxford
University, St. John's College (2001).

\bibitem{esr2} T. Nojiri, unpublished.


\bibitem{kato2}
R. Kato, N.    Tajima, M. Tamura, and J-I. Yamaura,
 Phys. Rev. B \textbf{66}, 020508 (2002).

\bibitem{tohno}
T. Miyazaki and T. Ohno,
 Phys. Rev. B \textbf{59}, 5269   (1999).

\bibitem{komatsu}
T. Komatsu, N. Matsukawa, T. Inoue, and G. Saito,
 J. Phys. Soc. Jpn. \textbf{65}, 1340 (1996).

\bibitem{merino3}
One should be cautious about taking H\"uckel and extended H\"uckel
values as definitive. Based on comparison with results from the
local density approximation it has been argued [J. Merino and R.
H. McKenzie, Phys. Rev. B {\bf 62}, 2416    (2000)] that the
H\"uckel methods tend to systematically underestimate the hopping
integrals.

\bibitem{girlando}
G. Visentini, M. Masino, C. Bellitto, and A. Girlando
Phys. Rev. B {\bf 58}, 9460 (1998).

\bibitem{qcref}
A. Fortunelli and A. Painelli,
Phys. Rev. B {\bf 55}, 16088 (1997).


\bibitem{nakamura}
T. Nakamura, T. Takahashi, S. Aonuma, and R. Kato, J. Mater. Chem.
\textbf{11}, 2159 (2001).

\bibitem{macdonald}
A.H. MacDonald, S.M. Girvin, and D. Yoshioka, Phys. Rev. B {\bf
41}, 2565 (1990); {\bf 37}, 9753 (1988).

\bibitem{misguich}
G. Misguich, C. Lhuillier, B. Bernu, and C. Waldtmann
 Phys. Rev. B \textbf{60}, 1064 (1999).

\bibitem{masutomi}
R. Masutomi, Y. Karaki, and H. Ishimoto,
 Phys. Rev. Lett. \textbf{92}, 025301 (2004).

\bibitem{ming}
 W. LiMing,  G. Misguich, P. Sindzingre, and C. Lhuillier,
 Phys. Rev. B {\bf 62}, 6372 (2000).

\bibitem{sc-dmit}
R. Kato, Y. Kashimura, S. Aonuma, N. Hanasaki, and H. Tajima,
Solid State Communications \textbf{105}, 561 (1998); J. I.
Yamaura, A. Nakao, and R. Kato,
 J. Phys. Soc. Jpn. \textbf{73}, 976  (2004).


\bibitem{capone}
 M. Capone, L. Capriotti, F. Becca, and  S. Caprara,
 Phys. Rev. B {\bf 63}, 085104 (2000).

\bibitem{schiffer}
P. Schiffer and I. Daruka,
 Phys. Rev. B {\bf 56}, 13712 (1997).

\bibitem{huber}
A. J. Garcia-Adeva and D. L. Huber,
 Phys. Rev. B {\bf 63}, 174433 (2001).

\bibitem{moessner3}
R. Moessner and A. J. Berlinsky,
 Phys. Rev. Lett. \textbf{83}, 3293   (1999).

\bibitem{merino2}
J. Merino and R. H. McKenzie,
 Phys. Rev. B {\bf 61}, 7996 (2000).

\bibitem{bernu2}
B. Bernu and G. Misguich,
 Phys. Rev. B {\bf 63}, 134409 (2001).

\end{thebibliography}


\newpage
\widetext

\begin{table}
\caption{Series coefficients for the high-temperature expansions of the
uniform susceptibility
$\overline{\chi}$ in Eq. (\ref{eq_chi}).
 Nonzero coefficients $c_{m,n}$
up to order $n=10$ are listed.}\label{tabht}
\begin{tabular}{|rr|rr|rr|rr|}
\hline
\hline
\multicolumn{1}{|c}{($m,n$)}  &\multicolumn{1}{c|}{$c_{m,n}$}
&\multicolumn{1}{c}{($m,n$)} &\multicolumn{1}{c|}{$c_{m,n}$}
&\multicolumn{1}{c}{($m,n$)} &\multicolumn{1}{c|}{$c_{m,n}$}
&\multicolumn{1}{c}{($m,n$)} &\multicolumn{1}{c|}{$c_{m,n}$} \\
\hline
  ( 0, 0) &            1 &  ( 3, 5) &        -7680 &  ( 7, 7) &        20480 &  ( 7, 9) &   -129328128  \\
  ( 0, 1) &           -4 &  ( 4, 5) &         1920 &  ( 0, 8) &      4205056 &  ( 8, 9) &   -159694848  \\
  ( 1, 1) &           -2 &  ( 5, 5) &         -672 &  ( 1, 8) &    -58877952 &  ( 9, 9) &     19133440  \\
  ( 0, 2) &           16 &  ( 0, 6) &        23488 &  ( 2, 8) &    110985216 &  ( 0,10) &  -2574439424  \\
  ( 1, 2) &           32 &  ( 1, 6) &       293376 &  ( 3, 8) &      -501760 &  ( 1,10) &  52032471040  \\
  ( 0, 3) &          -64 &  ( 2, 6) &       111552 &  ( 4, 8) &    101972480 &  ( 2,10) &   -735774720  \\
  ( 1, 3) &         -264 &  ( 3, 6) &       411392 &  ( 5, 8) &    -84013056 &  ( 3,10) & -29924454400  \\
  ( 2, 3) &          -96 &  ( 4, 6) &      -115968 &  ( 6, 8) &     29817856 &  ( 4,10) &  15318384640  \\
  ( 3, 3) &           16 &  ( 5, 6) &        70656 &  ( 7, 8) &    -15618048 &  ( 5,10) &  38033190912  \\
  ( 0, 4) &          416 &  ( 6, 6) &       -12768 &  ( 8, 8) &      2923776 &  ( 6,10) & -40192143360  \\
  ( 1, 4) &         1216 &  ( 0, 7) &       207616 &  ( 0, 9) &   -198295552 &  ( 7,10) &  48646737920  \\
  ( 2, 4) &         2400 &  ( 1, 7) &     -1766016 &  ( 1, 9) &   -571327488 &  ( 8,10) & -13533921280  \\
  ( 3, 4) &         -512 &  ( 2, 7) &     -7739648 &  ( 2, 9) &   3934844928 &  ( 9,10) &   4594278400  \\
  ( 4, 4) &           80 &  ( 3, 7) &     -1804992 &  ( 3, 9) &  -4115195904 &  (10,10) &   -869608960  \\
  ( 0, 5) &        -4544 &  ( 4, 7) &     -3373440 &  ( 4, 9) &   3772164096 &    &  \\
  ( 1, 5) &       -10880 &  ( 5, 7) &       689920 &  ( 5, 9) &  -1888413696 &    &  \\
  ( 2, 5) &       -20480 &  ( 6, 7) &       120064 &  ( 6, 9) &   1134317568 &    &  \\
\hline
\hline
\end{tabular}
\end{table}

\end{document}